\documentclass[alpha-refs]{wiley-article}
\usepackage{color}
\usepackage{ulem}

\usepackage{siunitx}
\usepackage{graphicx}
\usepackage{amsfonts} % added for equations
\usepackage{commath}
\usepackage{dsfont}
\usepackage{multirow}
\usepackage{lineno}
\usepackage{url}
\usepackage{colortbl} % For background color
\usepackage{xcolor}   % To define colors

\papertype{Original Article}
\title{A Deep-Learning-Driven Optimization-Based Inverse Solver for Accelerating the Marchenko Method}

\author[1]{Ning Wang}
\author[1]{Tariq Alkhalifah}

\affil[1]{Physical Science and Engineering Division, King Abdullah University of Science and Technology, Thuwal, Makkah Province, 23955-6900, Saudi Arabia}

\corraddress{Ning Wang, Physical Sciences and Engineering, King Abdullah University of Science and Technology, Thuwal, Makkah Province, 23955-6900, Saudi Arabia}
\corremail{ning.wang.2@kaust.edu.sa}

\fundinginfo{KAUST}

\runningauthor{Wang and Alkhalifah}

\begin{document}

\maketitle

\begin{abstract}
The Marchenko method is a powerful tool for reconstructing full-wavefield Green's functions using surface-recorded seismic data. These Green's functions can then be utilized to produce subsurface images that are not affected by artifacts caused by internal multiples. Despite its advantages, the method is computationally demanding, primarily due to the iterative nature of estimating the focusing functions, which links the Green’s functions to the surface reflection response. To address this limitation, an optimization-based solver is proposed to estimate focusing functions in an efficient way. This is achieved by training a network to approximate the forward modeling problem on a small subset of pre-computed focusing function pairs, mapping final up-going focusing functions obtained via the conventional iterative scheme to their initial estimates. Once trained, the network is fixed and used as the forward operator within the Marchenko framework. For a given target location, an input is initialized and iteratively updated through backpropagation to minimize the mismatch between the output of the fixed network and the known initial up-going focusing function. The resulting estimate is then used to compute the corresponding down-going focusing function and the full Green's functions based on the Marchenko equations. This strategy significantly reduces the computational cost compared to the traditional Marchenko method based on conventional iterative scheme. Tests on a synthetic model, using only 0.8\% of the total imaging points for training, show that the proposed approach accelerates the imaging process while maintaining relatively good imaging results, which is better than reverse time migration. Application to the Volve field data further demonstrates the method’s robustness and practicality, highlighting its potential for efficient, large-scale seismic imaging.

% Please include a maximum of seven keywords
\end{abstract}

%%%%%%%%%%%%%%%%%%%%%%%%%%%%
\section{Introduction}
The Marchenko method \citep{Broggini2012, Wapenaar2014, Broggini2014, vanderNeut2015, Wapenaar2017} is used to reconstruct up- and down-going Green’s functions between surface receivers and virtual sources (also called focal points). These Green’s functions include both primary reflections and internal multiples, making them valuable for artifact-free imaging. This reconstruction is achieved by solving the Marchenko equations for the up- and down-going focusing functions, using surface reflection data and a smooth migration velocity model. Once the focusing functions are obtained, they serve as a link between the surface reflection response and the Green’s functions (from focal points to surface receivers), allowing for the retrieval of the latter. However, the estimation of the focusing functions is computationally intensive, as it needs to be repeated for each imaging point and typically involves solving a truncated Neumann series~\citep{vanderNeut2015} or applying iterative schemes such as least squares with QR factorization (LSQR)~\citep{Ravasi2021}. Although the Marchenko method has been successfully applied to field data in various geological settings \citep{Ravasi2015, Ravasi2016, Jia2017, Staring2021}, its high computational cost limits its broader adoption. To make the Marchenko method more practical for large-scale seismic imaging, it is essential to develop computational strategies that significantly reduce the cost of estimating the focusing function while preserving acceptable accuracy.

In recent years, deep learning \citep[DL --][]{Geoffrey2006, LeCun2015} has emerged as an effective tool to overcome limitations in computational speed and cost, especially for large inverse problems. Among the various DL architectures, convolutional neural networks \citep[CNNs]{Lecun1998, Browne2003, Goodfellow2016} have shown exceptional ability to extract and process complex patterns from image. Building on the foundation of CNNs, the U-Net architecture, which is characterized by its symmetric encoder-decoder structure, was originally developed for biomedical image segmentation \citep{Ronneberger2015}. Due to its effectiveness in capturing both local and global features, U-Net has gained popularity in seismic applications, where it has been successfully used to learn intricate mappings~\citep{Birnie2022, Alfarhan2024}. Leveraging this capability, \cite{Wang2025} proposed a deep learning-based method using U-Net to predict up-going focusing functions using a U-Net architecture, followed by computing down-going focusing functions through established physical relationships. Although this data-driven approach enables rapid predictions for selected focal points, its application to large-scale imaging remains limited by the reduced generalization capacity of fully supervised models.

To overcome this limitation, physics-constrained learning frameworks have been explored. One such approach is Physics-Informed Neural Networks (PINNs), which incorporate physical constraints, such as partial differential equations, directly into the loss function during training \citep{Raissi2019, Song2022, Zhang2023}. Building on this idea, optimization-based inverse solvers take a complementary approach. Rather than training the network to learn a full solution mapping, these methods use a pre-trained neural network as a fixed, differentiable forward operator. During inference, the input to the network is iteratively optimized via backpropagation (BP) to satisfy a physics-based loss function. In this formulation, the physical consistency is enforced through the loss function used to guide the input optimization, enabling greater flexibility and improved generalization without retraining. This concept was first demonstrated by \citet{Loper2014} through OpenDR, a differentiable renderer that introduced gradient-based optimization. More recently, similar strategies have been successfully applied to accelerate full waveform inversion using fixed neural operators and automatic differentiation \citep{Zou2024}. These optimization-based inverse solvers offer a improved generalization and physics-constrained adaptability without retraining the network.

Leveraging the demonstrated advantages of optimization-based inverse solvers, such as their efficiency in updating inputs via automatic differentiation (AD) and their capability to incorporate physics-based constraints, we propose a novel approach to accelerate Marchenko imaging. In this approach, a neural network is first trained on a small subset of up-going focusing functions randomly selected from the target imaging area. The network learns to map final up-going focusing functions (computed using LSQR) to their corresponding initial estimates (computed using a smooth background velocity model). Once trained, the network is fixed and treated as a differentiable forward operator within an optimization loop. For each new focal point, an initial guess of the up-going focusing function is iteratively updated by minimizing a loss function that quantifies the misfit between the network’s output and the known initial focusing function. Gradients are computed via AD through the fixed network to iteratively refine the input. After the up-going focusing function is obtained, the corresponding down-going focusing function is then directly computed from the surface reflection data using the Marchenko equations. The proposed method only requires few training and validation data that is derived from the target imaging area, making it highly adaptable and practical for new seismic tasks or domains. By replacing the traditional, computationally expensive inversion with a cost-effective AD-based optimization, the proposed approach addresses the computational challenges that have limited the broader application of the Marchenko method.
The effectiveness of the proposed method is validated on both synthetic and field data. For the synthetic model, the inverted up-going and derived down-going focusing functions, along with the resulting seismic images, closely match those obtained from the conventional iterative scheme. Further validation is provided through application to the Volve field dataset, where the results show minimal signal leakage compared to the conventional Marchenko method, highlighting the method’s practical reliability and adaptability. By replacing the traditional iterative inversion with an optimization-based solver, the proposed approach significantly reduces computational time while maintaining imaging quality. This makes it a robust and scalable solution for applying the Marchenko method in large-scale seismic imaging tasks.

\section{Theory}
In this section, we first review the formulations behind Marchenko redatuming. Then, we describe our neural network optimization-based framework for inverting focusing functions corresponding to image points. The network architecture and pre-training are described afterwards. Finally, we deep dive into the inversion process and discuss the ingredients for proper convergence to an acceptable focusing function.

\subsection{Marchenko Redatuming}
The Marchenko method~\citep{Wapenaar2014} provides a framework for retrieving wavefields based on a physical model of wave propagation in a heterogeneous, lossless medium. The acquisition setup includes a transparent surface $\Lambda_{S}$ where sources and receivers are co-located at positions $\textbf{x}_S$ and $\textbf{x}_R$, and a subsurface level $\Lambda_{F}$ where virtual sources (or focal points) are placed. Within this setup, the Marchenko equations relate the up- and down-going components of two key wavefields: the Green’s functions, which describe the response propagating from surface receivers $\textbf{x}_R$ to a focal point $\textbf{x}_{F}$ at depth $\Lambda_{F}$ ($g^{-}(\textbf{x}_{R},\textbf{x}_{F})$ and $g^{+}(\textbf{x}_{R},\textbf{x}_{F})$), and the focusing functions, which propagate from a focal point to the surface ($f^{-}(\textbf{x}_{S},\textbf{x}_{F})$ and $f^{+}(\textbf{x}_{S},\textbf{x}_{F})$). These wavefields are linked through the surface reflection response ${R}(\textbf{x}_{R},\textbf{x}_S)$, expressed as:
\begin{equation}
\label{eq:Marchenko}
\begin{aligned}
g^{-}(\textbf{x}_{R},\textbf{x}_{F}) &= \int _{\Lambda_S}R(\textbf{x}_{R},\textbf{x}_S)f^{+}(\textbf{x}_S,\textbf{x}_{F})d\textbf{x}_S - f^{-}(\textbf{x}_{R},\textbf{x}_{F}), \\
-g^{+*}(\textbf{x}_{R},\textbf{x}_{F}) &= \int _{\Lambda _S}R^{*}(\textbf{x}_{R},\textbf{x}_S)f^{-}(\textbf{x}_S,\textbf{x}_{F})d\textbf{x}_S - f^{+}(\textbf{x}_{R},\textbf{x}_{F}),
\end{aligned}
\end{equation}
where superscripts are used to define the up-going (-) and down-going (+) components, and the asterisk $*$ indicates complex conjugation in the frequency domain, which corresponds to time reversal in the time domain.

To solve the Marchenko equations for the focusing functions, a windowing operator $\Theta_{g}$ is applied on both sides of equation \ref{eq:Marchenko}. This operator removes wavefield contributions that arrive after the traveltime $t_{d}(\textbf{x}_{R},\textbf{x}_{F})-\epsilon$ and before its time-reversed counterpart $-(t_{d}(\textbf{x}_{R},\textbf{x}_{F})-\epsilon)$. Here, $t_{d}(\textbf{x}_{R},\textbf{x}_{F})$ denotes the traveltime of the direct wave from the focal point $\textbf{x}_{F}$ to surface receivers $\textbf{x}_{R}$, and $\epsilon$ represents half the wavelet duration. Since the Green’s functions happen only after the traveltime of the direct arrival and before the negative traveltime, applying this window operator $\Theta_{g}$ eliminates their contribution, effectively reducing them to zero ($\Theta_{g}{g^{-}(\textbf{x}_{R},\textbf{x}_{F})}={0}$ and $\Theta_{g}{g^{+}(\textbf{x}_{R},\textbf{x}_{F})}={0}$). This simplification yields a modified version of the Marchenko equations that depends only on the focusing functions and the reflection response:
\begin{equation}
\label{eq:Marchenko_ff}
\begin{aligned}
-\Theta_{g}\int _{\Lambda _{S}}R(\textbf{x}_{R},\textbf{x}_S)f_{d}^{+}(\textbf{x}_S,\textbf{x}_{F})d\textbf{x}_S &= \Theta_{g}\int _{\Lambda _{S}}R(\textbf{x}_{R},\textbf{x}_S)f^{+}_{m}(\textbf{x}_S,\textbf{x}_{F})d\textbf{x}_S - f^{-}(\textbf{x}_{R},\textbf{x}_{F}), \\
0 &= \Theta_{g}\int _{\Lambda _{S}}R^{*}(\textbf{x}_{R},\textbf{x}_S)f^{-}(\textbf{x}_S,\textbf{x}_{F})d\textbf{x}_S - f^{+}_{m}(\textbf{x}_{R},\textbf{x}_{F}),
\end{aligned}
\end{equation}
which can be expressed in the following compact matrix-vector notation as:
\begin{equation}
\label{eq:Marchenko_ff_compact}
\begin{bmatrix}
    \Theta_{g}\mathbf{Rf_{d}^{+}} \\
    \mathbf{0}
\end{bmatrix}
=
\begin{bmatrix}
    \mathbf{I} & -\Theta_{g}\mathbf{R} \\
    -\Theta_{g}\mathbf{R^{*}} & \mathbf{I} 
\end{bmatrix}
\begin{bmatrix}
    \mathbf{f^{-}} \\
    \mathbf{f^{+}_{m}}  
\end{bmatrix}
.
\end{equation}
In this formulation, the down-going focusing function $\mathbf{f^{+}}$ is decomposed into two parts: the direct arrival $\mathbf{f_{d}^{+}}$, which can be computed using a smooth migration velocity model (e.g. via Eikonal or wave equations), and its coda $\mathbf{f_{m}^{+}}$. With only two unknowns remaining, the simplified equation \ref{eq:Marchenko_ff} can be solved iteratively. In this study, we employ the LSQR algorithm \citep{Paige1982} to recover these focusing functions. Once retrieved, the focusing functions can be substituted back into the original Marchenko equations (equation \ref{eq:Marchenko}) to compute the full up- and down-going Green’s functions. It is important to note that if the up-going focusing function $\mathbf{f^{-}}$ is available, the coda of the down-going focusing function, $\mathbf{f_{m}^{+}}$, can be directly calculated using:
\begin{equation}
\label{eq:ff}
\mathbf{f^{+}_{m}} = \Theta_{g}\mathbf{R^{*}} \mathbf{f^{-}}
.
\end{equation}
The complete down-going focusing function $\mathbf{f^{+}}$ is then reconstructed by summing the direct and coda components, yielding $\mathbf{f^{+}} = \mathbf{f^{+}_{d}} + \mathbf{f^{+}_{m}}$.

\subsection{Optimization-based inverse solver Framework}
To overcome the high computational cost associated with solving equation \ref{eq:Marchenko_ff_compact} via conventional iterative methods, we propose to adopt an optimization-based framework that estimates focusing functions by treating a pre-trained neural network $\mathcal{N}_\theta$ as a differentiable forward operator. This approach consists of two main stages: network training and inversion.

\textbf{Stage 1: Network Training}

\begin{figure*}[!t]
\centering
\includegraphics[width=0.85\textwidth]{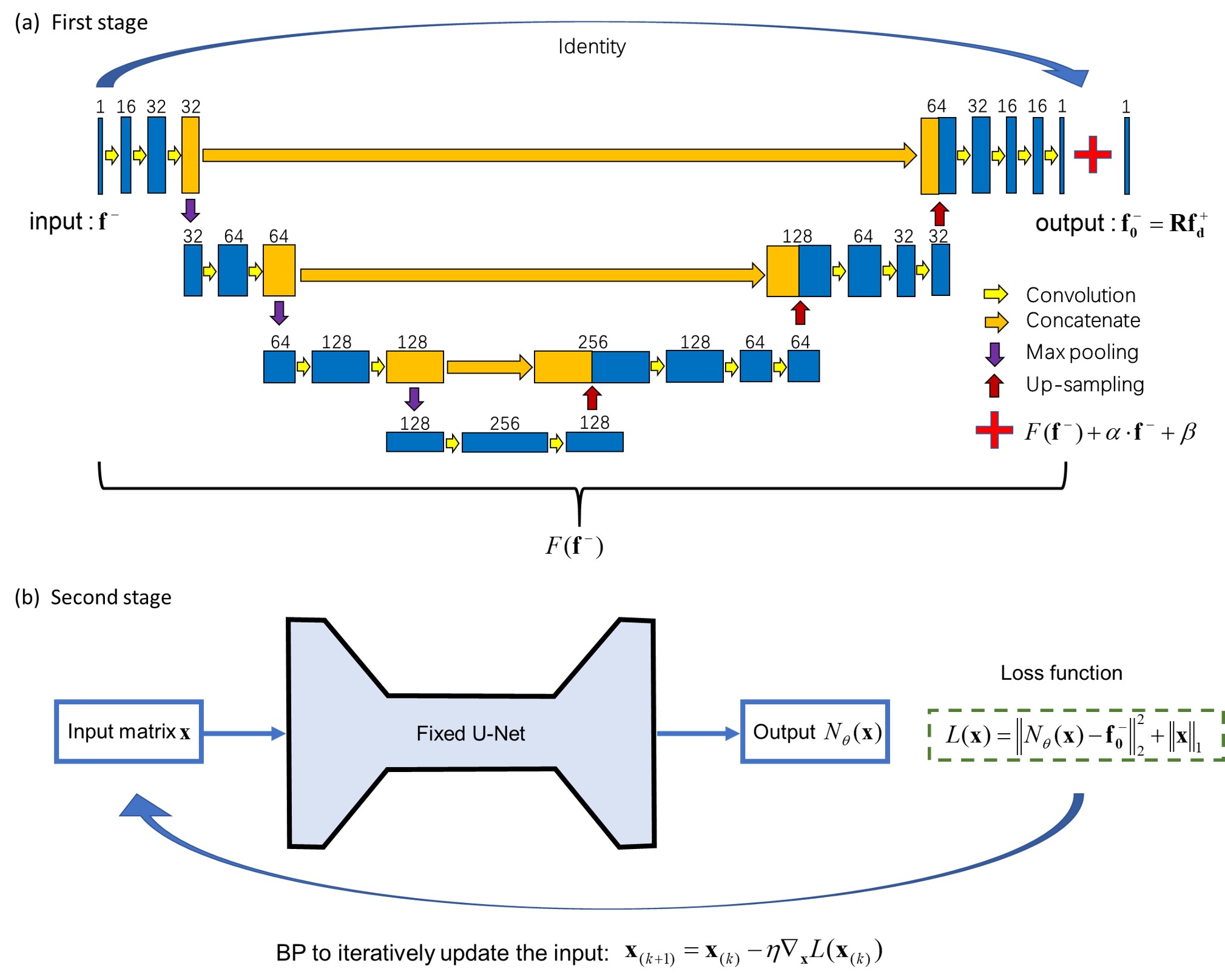}
\caption{Two-stage framework of the proposed optimization-based inverse solver. (a) Network Training: a U-Net learns the mapping from final to initial up-going focusing functions. (b) Inversion: the trained network is fixed, and inputs are optimized via BP to recover up-going focusing functions at new focal points.}
\label{figs:U-Net}
\end{figure*}

A small set of focal points is randomly selected from the target imaging area, and their focusing functions are computed using LSQR solver. These pre-computed focusing functions are then used to train a neural network $\mathcal{N}_\theta$ that maps the final up-going focusing functions $\mathbf{f^{-}}$ to their corresponding initial approximation $\mathbf{f^{-}_0} = \mathbf{Rf_{d}^{+}}$, as shown in Figure \ref{figs:U-Net}a. Once trained, the network parameters $\theta$ are fixed, and the network is used as a differentiable forward operator in the inversion stage.

\textbf{Stage 2: Inversion}

For a new focal point where no pre-computed focusing function is available, an input matrix $\mathbf{x}_0$ is initialized and the corresponding initial up-going focusing function $\mathbf{f^{-}_0}$ is used as the optimization target. The objective is to determine an input $\mathbf{x}$ such that the output of trained network $\mathcal{N}\theta(\mathbf{x})$ closely matches the target $\mathbf{f}_0^{-}$. To achieve this, the following loss function $\mathcal{L}$ is minimized:
\begin{equation}
\label{eq:loss}
\mathcal{L}(\mathbf{x}) = \left\| \mathcal{N}_\theta(\mathbf{x}) - \mathbf{f^{-}_0} \right\|_2^2 + \lambda \left\| \mathbf{x} \right\|_1
,
\end{equation}
where the first term enforces data fidelity and the second term, an $\ell_1$ regularization, promotes sparsity in the input to suppress inversion noise. The optimization is carried out iteratively, updating $\mathbf{x}$ using the gradient of the loss function with respect to the input, which is computed via BP:
\begin{equation}
\label{eq:update}
\mathbf{x}_{(k+1)} = \mathbf{x}_{(k)} - \eta \, \nabla_{\mathbf{x}} \mathcal{L}(\mathbf{x}_{(k)})
,
\end{equation}
where $\eta$ is the learning rate and $k$ denotes the iteration number. This process continues until a convergence criterion is met or a fixed number of iterations is reached (Figure \ref{figs:U-Net}b), providing an optimized input $\mathbf{x}_{(k+1)}$, which corresponds to the estimated up-going focusing function $\hat{\mathbf{f}}^{-} = \mathbf{x}_{(k+1)}$. Then, the coda of the down-going focusing function, $\mathbf{f^{+}_{m}}$, is computed using equation \ref{eq:ff}. Adding the direct arrival yields the full down-going focusing function, $\mathbf{f}^{+} = \mathbf{f_{d}}^{+} + \mathbf{f_{m}}^{+}$. 
An Adam optimizer \citep{Kingma2014} is employed in both the training stage (to update the network parameters) and the inversion stage (to optimize the input $\mathbf{x}$).

\subsection{The network architecture and training}
The neural network used in this study follows the standard U-Net architecture \citep{Ronneberger2015}, consisting of a series of contracting blocks in the encoder and corresponding expanding blocks in the decoder, connected via skip connections, as shown in Figure \ref{figs:U-Net}a. To improve training stability and generalization, we apply Leaky ReLU activation functions with a negative slope of 0.2, along with batch normalization and 50\% dropout after each convolutional layer.
Given that the network input ($\mathbf{f}^{-}$) and target ($\mathbf{f}_0^{-}$) are kinematically similar, differing primarily in the presence of weak multiple-related events, a residual connection is introduced to better capture these subtle features. Specifically, a learnable scalar scale $\alpha$ and bias $\beta$ are applied to the input before computing the residual (Figure~\ref{figs:U-Net}a), enabling the network to account for amplitude differences between the input and target wavefields. The final output is expressed as $F(\mathbf{f}^{-}) + \alpha \cdot \mathbf{f}^{-} + \beta$, where $F(\cdot)$ denotes the output of the U-Net.
To further guide the learning process, a time mask is applied to restrict the loss computation to the physically meaningful portions of the wavefield, which occur only before the direct arrival or after its time-reversed counterpart.

\subsection{The inversion}
It is important to emphasize that during the inversion stage, the choice of initial input $\mathbf{x}_0$ and the learning rate $\eta$, which controls the step size in the optimization loop, play a critical role in determining both the convergence speed and the accuracy of the reconstructed up-going focusing functions. Initializing $\mathbf{x}_0$ with a random matrix often leads to slow convergence due to the lack of prior structural information. On the other hand, directly using the initial up-going focusing function $ \mathbf{f}^{-}_0$ as input may result in poor handling of weaker or subtle events.

To balance convergence efficiency and reconstruction fidelity, we adopt an initialization strategy that expands outward from focal points with known final up-going focusing functions (training/validation points). For each known point, we select its closest unsolved test locations and form the initial guess $\mathbf{x}_0$ by spatially shifting the known focusing function according to the relative offset between the known and test positions. The magnitude and direction of the shift are determined by the relative spatial positions of the target and the known points. The network input $\mathbf{x}$ is then iteratively optimized via BP to minimize the loss defined in equation~\eqref{eq:loss}, yielding a refined up-going focusing function $\hat{\mathbf{f}}^{-}$ at the new location. Newly recovered solutions are added to the pool and used to initialize next ring of neighboring test points. This process is repeated iteratively, producing a region-growing process for the whole target imaging area. This strategy provides a physically meaningful initialization, which accelerates and stabilizes convergence and improves the recovery of weak or subtle events that are otherwise difficult to update.

In addition, the learning rate $\eta$ also affects inversion performance. If $\eta$ is too small, convergence is slow and computational cost increases; if too large, the inverse procedure can become unstable. In this work, we adopt a slightly larger learning rate to speed convergence, accepting a modest increase in noise as a trade-off. The learning rate is selected using the training data to ensure stable, efficient convergence and then applied unchanged during inversion of the test data.

In summary, the proposed method comprises two main stages. In the training stage, a small subset of focal points within the imaging area is randomly selected, and their focusing functions are computed using the conventional iterative scheme. These are used to train a neural network that learns the forward mapping from final up-going focusing functions to their corresponding initial approximations. In the inversion stage, the trained network is fixed and employed as a differentiable forward operator. For each unseen focal point, the input is iteratively optimized to minimize the loss function, resulting in an estimated up-going focusing function. The corresponding down-going focusing function, $\mathbf{f}^{+}$, is then recovered using equation~\ref{eq:ff} and the direct arrival $\mathbf{f}_{d}^{+}$. The complete set of focusing functions is subsequently substituted into equation~\ref{eq:Marchenko} to compute the up- and down-going Green’s functions for imaging. This optimization-based inverse workflow enables the estimation of focusing functions without solving the full Marchenko equations for every point, significantly reducing the computational cost associated with large-scale imaging. Moreover, it provides flexibility to incorporate regularization during inversion, which can enhance stability, suppress noise, and promote desirable properties such as sparsity or smoothness in the recovered wavefields.

\section{Numerical Examples}
In this section, we first test the proposed approach on synthetic data, which is helpful in evaluating its accuracy considering the potential speed grained compared to numerical methods. We then apply the approach on the Volve field dataset. 
\subsection{Synthetic example}
\begin{figure*}[!t]
\centering
\includegraphics[width=1\textwidth]{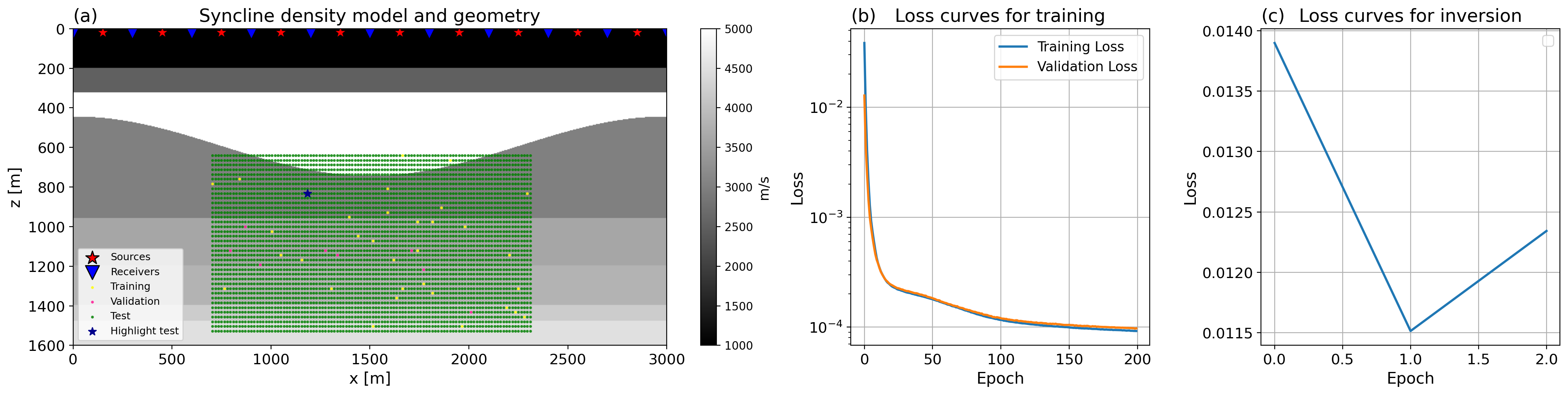}
\caption{(a) Syncline density model and geometry, with yellow and pink dots representing the random points where the focusing functions are calculated numerically and used for training and validation of our U-Net. (b) Training and validation loss curves for network training stage. (c) Loss curves for inversion stage.}
\label{figs:geometry_simple}
\end{figure*}

To assess the performance of the proposed approach, we apply it to a synthetic dataset generated using a constant-velocity (2400 m/s), variable-density model, as shown in Figure~\ref{figs:geometry_simple}a. The simulation excludes free-surface effects and includes 201 co-located source and receiver positions, where sources are indicated by red stars and receivers by blue triangles on the surface. In order to perform imaging, the imaging points are selected on a regular grid composed of 38 vertical lines, each spaced 24 m apart. Along each line, 108 points are sampled at 15 m vertical intervals, resulting in a total of 4104 imaging points covering the rectangular area of interest. In this experiment, 0.8\% of the imaging points are randomly selected for training (yellow points), and an additional 0.2\% are used for validation (pink points). The focusing functions for these points are computed using the LSQR method. The remaining 99\% of the points (pink points) serve as test data and, more importantly, are computed through the proposed framework in a far more efficient manner to complete the imaging region.
The U-Net architecture consists of five levels of down-sampling and up-sampling blocks, with channel widths set to 16, 32, 64, 128, and 256, respectively. The first training stage is performed with a batch size of 16 and a learning rate of 0.001. As shown in the training and validation convergence curves in Figure \ref{figs:geometry_simple}b, the training process is stopped after 200 epochs. After training, the network weights are fixed and used as a differentiable operator in the inversion process.

\begin{figure*}[!t]
\centering
\includegraphics[width=1\textwidth]{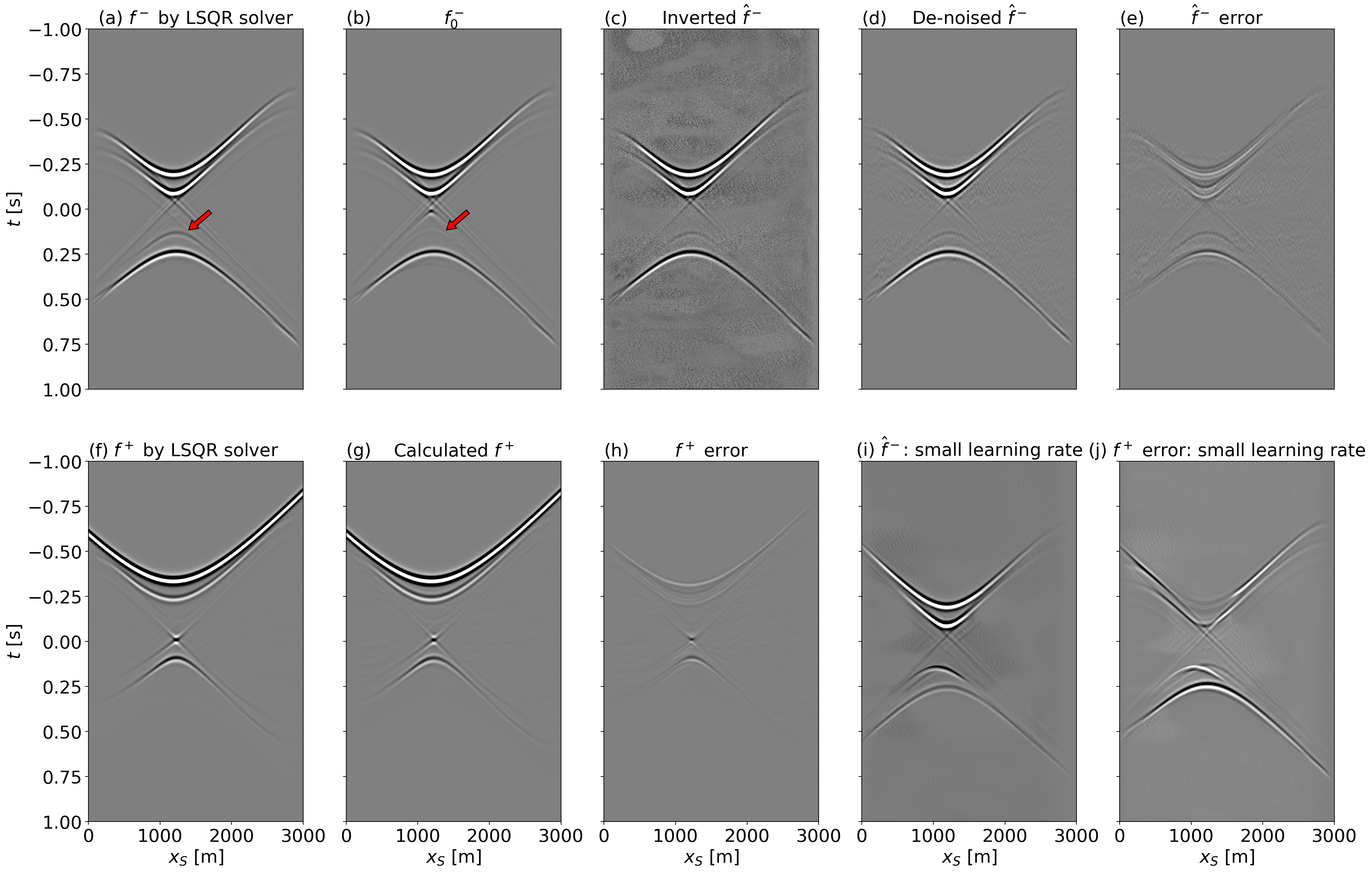}
\caption{Comparison of focusing functions obtained using different methods. (a) Up-going focusing function computed with the LSQR solver (reference), (b) initial up-going focusing function derived from the direct arrival and reflection response, (c) up-going focusing function recovered using the proposed optimization-based inverse solver, and (d) its denoised version after applying a time window and band-pass filter. (e) Difference between the up-going focusing functions from the LSQR solver and proposed methods. Down-going focusing functions obtained using (f) the LSQR solver and (g) the denoised up-going focusing function via Equation~\ref{eq:ff}. (h) Difference between the down-going focusing functions from the two methods. (i) Up-going focusing function recovered using a smaller learning rate (0.001) and (j) its difference with the LSQR result.}
\label{figs:f1_inv_simple}
\end{figure*}

During the inversion stage, up-going focusing functions at the test points are recovered by iteratively optimizing the initial guess via BP through the fixed neural network. In this experiment, the learning rate is set to 0.1, and the inversion is terminated after two iterations. The corresponding convergence curve is shown in Figure~\ref{figs:geometry_simple}c. To illustrate the effectiveness of the proposed method, we present results for a focal point selected from the test set, marked by a blue star in Figure~\ref{figs:geometry_simple}a. This point represents a particularly challenging case, as it lies in a region with few nearby training or validation samples. 
The reference up-going focusing function $\mathbf{f}^{-}$, computed using the LSQR solver, is shown in Figure~\ref{figs:f1_inv_simple}a. The initial up-going focusing function $\mathbf{f}_0^{-}$, derived using the direct arrival and reflection response, is shown in Figure~\ref{figs:f1_inv_simple}b. As highlighted by the red arrows, several weak events are present in the final solution but absent in the initial estimate.
The inverted result after two iterations is shown in Figure~\ref{figs:f1_inv_simple}c, where the primary events are well recovered, although some noise is introduced due to the relatively high learning rate. To mitigate this noise, a time window operator is applied to isolate the valid region, while a band-pass filter (5–40 Hz) is used to suppress low- and high-frequency noise. The resulting denoised up-going focusing function is shown in Figure~\ref{figs:f1_inv_simple}d, which closely matches the LSQR-based reference (Figure~\ref{figs:f1_inv_simple}a), with only minor signal leakage visible in the difference plot (Figure~\ref{figs:f1_inv_simple}e). The down-going focusing function reconstructed using the inverted up-going component (Figure~\ref{figs:f1_inv_simple}g) also aligns well with the LSQR result (Figure~\ref{figs:f1_inv_simple}f), and the residual between the two (Figure~\ref{figs:f1_inv_simple}h) confirms minimal discrepancy.

To demonstrate the importance of using a relatively large learning rate, we repeat the inversion using a smaller learning rate of 0.001 and extend the number of iterations to 200. The resulting up-going focusing function for the same test point is shown in Figure~\ref{figs:f1_inv_simple}i, with the corresponding error displayed in Figure~\ref{figs:f1_inv_simple}j. The error plot reveals a noticeable shift, indicating that the lower learning rate not only slows down the inversion process but also limits the network's ability to escape local minima and correct shifts present in the initial input.

\begin{figure*}[!t]
\centering
\includegraphics[width=0.92\textwidth]{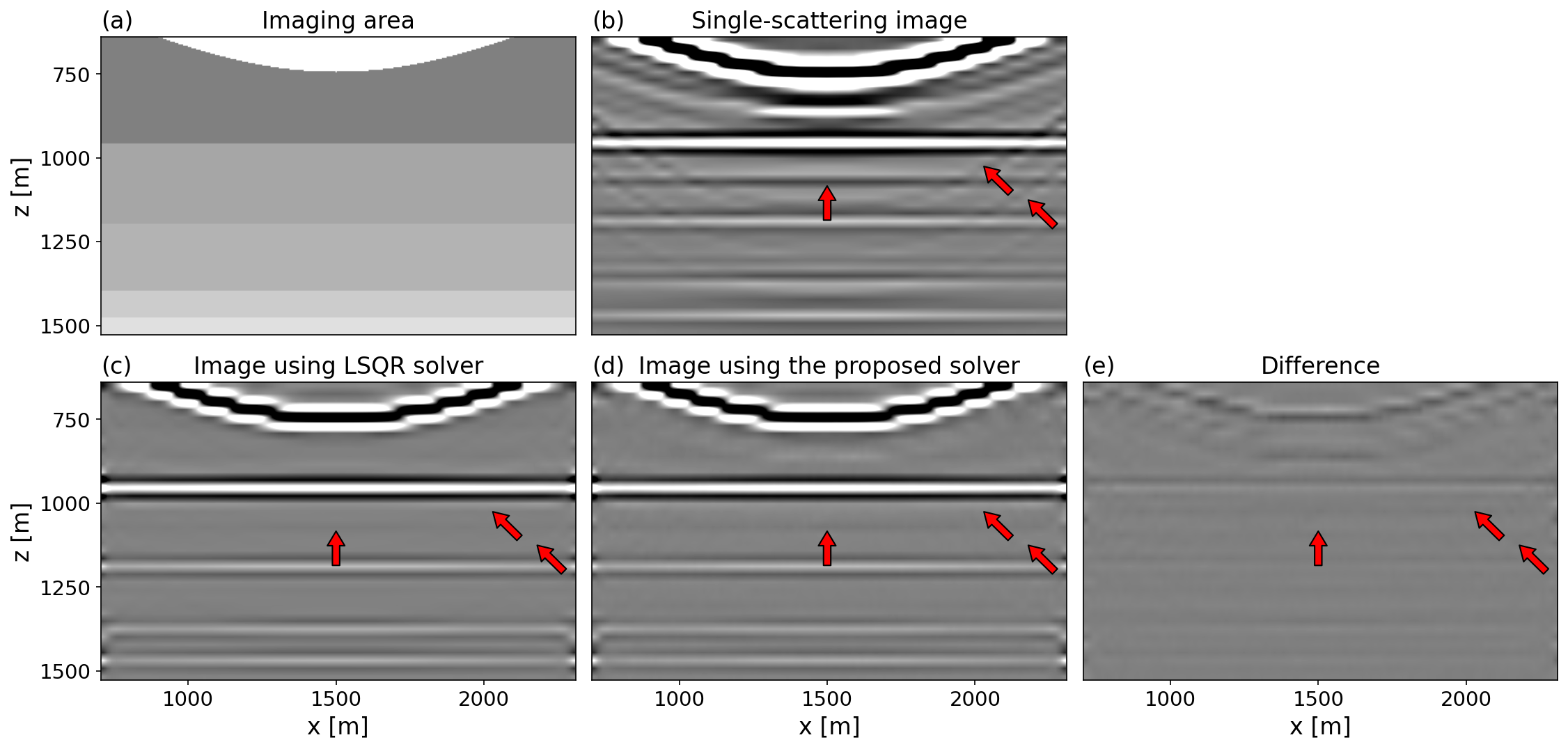}
\caption{(a) Target area used for imaging. (b) Image produced using the single-scattering method. (c) Image obtained using the Marchenko method with focusing functions computed via the LSQR solver. (d) Image obtained using the Marchenko method with focusing functions obtained by the proposed optimization-based inverse solver.}
\label{figs:Images_simple}
\end{figure*}

Next, we perform imaging using Green’s functions computed from the focusing functions obtained by both the conventional and proposed methods for the imaging area shown in Figure~\ref{figs:Images_simple}a. For comparison, an image produced using a single-scattering imaging formulation is presented in Figure \ref{figs:Images_simple}b. Figures~\ref{figs:Images_simple}c and \ref{figs:Images_simple}d show the resulting images using Green’s functions computed via the LSQR solver and the optimization-based inverse solver, respectively. The two images show excellent match with the true density model, with minimal differences, as highlighted in Figure \ref{figs:Images_simple}e. In contrast, the single-scattering image contains noticeable artifacts related to internal multiples, as indicated by the red arrows, which degrade the overall quality. These results confirm the effectiveness of the proposed method in accurately reconstructing focusing functions and handling internal multiples during imaging.

Finally, we evaluate the computational efficiency of the proposed method in comparison to the conventional Marchenko method. Using the LSQR solver, the computation time per imaging point is approximately 10.63 seconds, leading to a total processing time of about 12.12 hours for the entire imaging area. In contrast, the proposed optimization-based inverse solver achieves a substantial reduction in runtime, completing the process in approximately 36.37 minutes.
This total includes 7.27 minutes for computing the focusing functions of 1\% of the training and validation points, 7.23 minutes for network training, 7.24 minutes for performing inversion using the trained network, 10.41 minutes for post-processing with a band-pass filter and windowing operator, and 4.22 minutes to compute the down-going focusing functions. This results in an approximately 20-fold improvement in efficiency compared to the conventional method.
Despite the significant speed-up, the proposed framework maintains imaging quality that is comparable to that of the LSQR-based Marchenko method. This highlights the potential of the optimization-based inverse solver to reduce the computational cost of the Marchenko method while maintaining comparable image quality to the conventional one.

\subsection{Field Example}
\begin{figure*}[!t]
\centering
\includegraphics[width=1\textwidth]{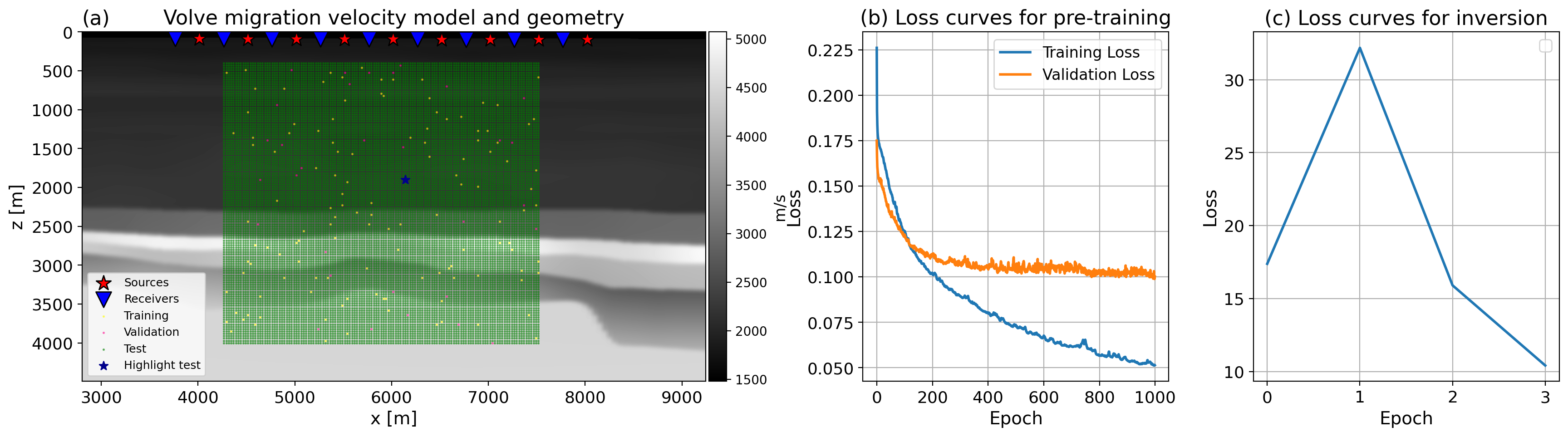}
\caption{(a) The 2D migration velocity model and acquisition geometry. (b) Training and validation loss curves for network training stage in the field example. (c) Loss curves for inversion stage in the field example.}
\label{figs:geometry_volve}
\end{figure*}

We further validate the proposed method on a field dataset acquired from the Volve field, located in the central North Sea, offshore Norway~\cite{Szydlik2007}. The application focuses on a two-dimensional (2D) acquisition geometry consisting of 180 ocean-bottom sensors spaced at 25 meters, paired with a nearby line of 110 sources spaced at 50 meters. A 2D velocity slice (Figure~\ref{figs:geometry_volve}a) is extracted from the 3D migration velocity model along the line nearest to the selected receivers to compute the direct arrival component $\mathbf{f}^{+}_{d}$. Imaging points are distributed across a 2D grid of 121 vertical lines spaced 30 meters apart. Each line contains 131 points sampled at 25-meter vertical intervals, resulting in a total of 15851 imaging points. From this set, 0.8\% of the points (marked in yellow) are randomly selected for training, and 0.2\% (in pink) are chosen for validation, with their corresponding focusing functions are computed using an LSQR-based solver. After training, the network parameters are fixed, and BP is applied to recover the up-going focusing functions for the remaining 99\% of imaging points (shown in green). The same U-Net architecture and hyper-parameters used in the synthetic experiment are adopted in the field example, with the exception of a reduced batch size of 8 to accommodate the field data volume.

To prepare the input for Marchenko redatuming, we must remove the surface-related multiples from the reflection response, and reorganize the data so we have co-located sources and receivers. We follow the workflow outlined in~\cite{Ravasi2016}, applying multi-dimensional deconvolution to construct the virtual reflection response $R(\mathbf{x}{R}, \mathbf{x}{R}')$. This response represents waves originating from virtual sources $\textbf{x}_{R}'$, positioned at the same spatial coordinates as the physical receivers, and being recorded at receiver positions $\textbf{x}_{R}$. It contains only primary reflections and internal multiples, and is also free of source signature effects.

The training phase is carried out over 1000 epochs, with convergence curves shown in Figure~\ref{figs:geometry_volve}b. Once trained, the network is fixed and used to estimate the up-going focusing functions at the test locations via iterative optimization through BP. A learning rate of 10 is used, and the inversion is terminated after three iterations. The convergence behavior during inversion is shown in Figure~\ref{figs:geometry_volve}c.

\begin{figure*}[!t]
\centering
\includegraphics[width=0.85\textwidth]{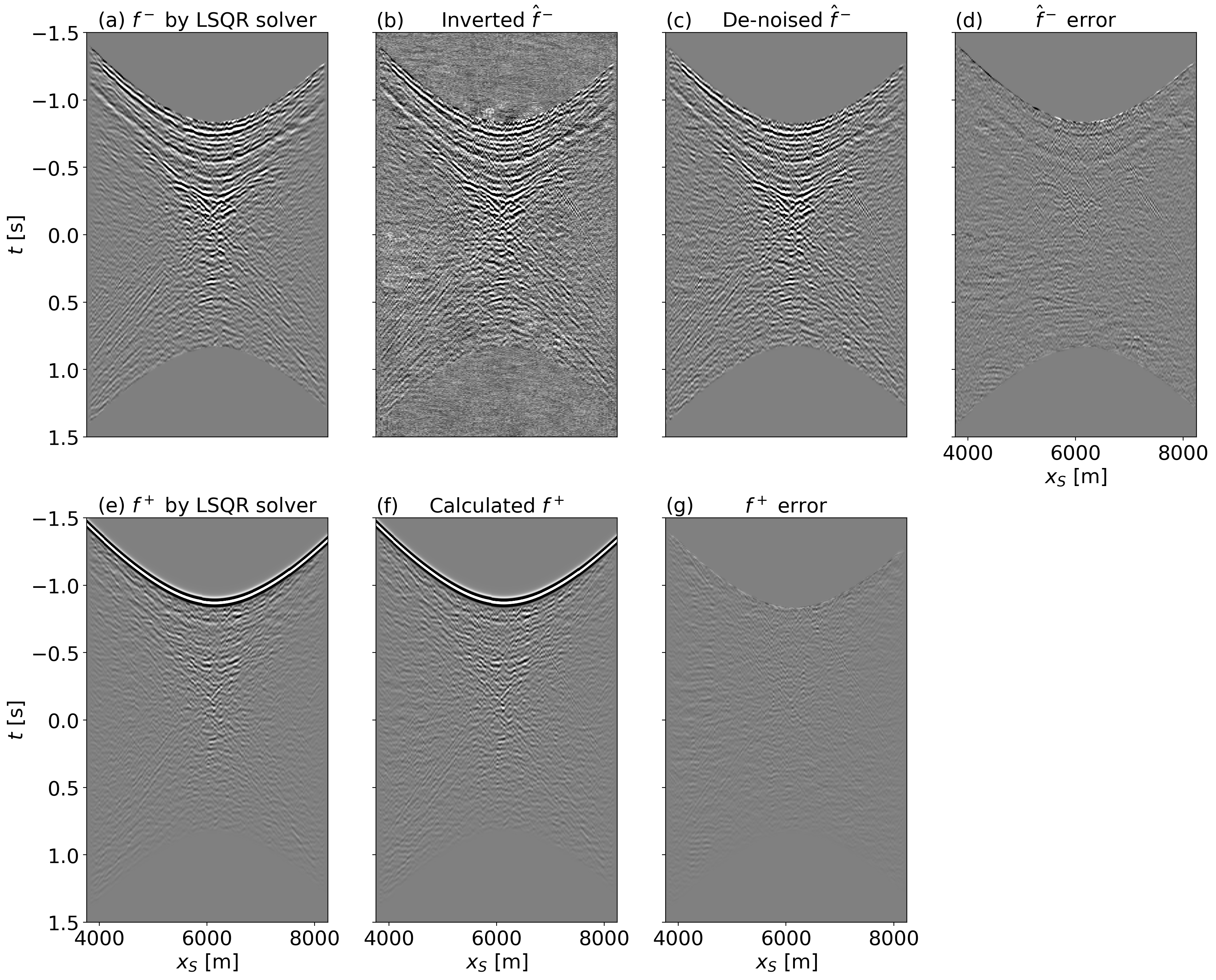}
\caption{Comparison of focusing functions obtained using different methods for Volve field data. (a) Up-going focusing function computed using the LSQR solver (reference), (b) up-going focusing function recovered by the proposed optimization-based inverse solver, (c) denoised version of (b) after applying a band-pass filter and a windowing operator. (d) Difference between (a) and (c). Down-going focusing function computed using (e) the LSQR solver and (f) the denoised up-going focusing function via Equation~\ref{eq:ff}. (g) Difference between (e) and (f).}
\label{figs:f1_inv_volve}
\end{figure*}

\begin{figure*}[!t]
\centering
\includegraphics[width=1\textwidth]{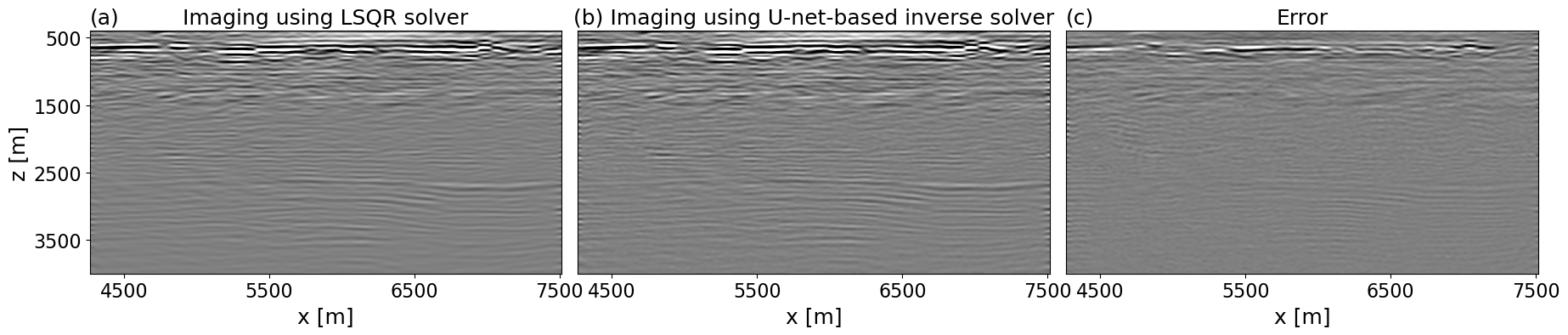}
\caption{Image obtained using focusing functions computed with (a) the LSQR solver and (b) the proposed optimization-based inverse solver, (c) Difference between the two images.}
\label{figs:imaging_volve}
\end{figure*}

The effectiveness of the proposed method is validated by showing the results at one test point indicated by the blue star as shown in Figure \ref{figs:geometry_volve}a. The up-going focusing function obtained using the LSQR solver is presented in Figure \ref{figs:f1_inv_volve}a, while the corresponding result from the optimization-based inverse solver is displayed in Figure~\ref{figs:f1_inv_volve}b. After applying a windowing operator and a band-pass filter (5–40 Hz), a denoised version of the predicted wavefield is obtained (Figure~\ref{figs:f1_inv_volve}c). The difference between the up-going focusing function estimated using the proposed method and the reference result, shown in Figure~\ref{figs:f1_inv_volve}d, confirms a close match with only minor discrepancies. The down-going focusing functions obtained from both methods are presented in Figures~\ref{figs:f1_inv_volve}e and \ref{figs:f1_inv_volve}f, respectively, showing strong consistency. Their difference plot (Figure~\ref{figs:f1_inv_volve}g) again indicates only minimal signal leakage. The final imaging results derived from the Marchenko method using the LSQR-based and optimization-based focusing functions are shown in Figures~\ref{figs:imaging_volve}a and \ref{figs:imaging_volve}b, respectively. The two images exhibit a high degree of agreement, with only minor variations highlighted in the difference plot in Figure~\ref{figs:imaging_volve}c. These results demonstrate the effectiveness of the proposed optimization-based inverse solver in accurately reconstructing focusing functions and enabling reliable Marchenko imaging for field data applications.

Finally, we assess the computational performance of both methods for the field experiment. The LSQR-based approach requires approximately 26.05 seconds per imaging point, resulting in a total processing time of roughly 114.7 hours for all imaging points. In contrast, the proposed optimization-based inverse solver drastically accelerates the process, reducing the total computation time to approximately 7.43 hours, which includes 1.15 hours for computing the focusing functions of the training and validation set, 3.77 hours for training, 1.24 hours for the inversion of up-going focusing functions, 32.76 minutes for de-noising, and 43.66 minutes for computing the down-going focusing functions.
Overall, the proposed method achieves an approximately 15-fold speed-up while delivering imaging results that are nearly indistinguishable from those produced by the LSQR solver. The successful application to field data further demonstrates the robustness and scalability of the optimization-based inverse solver, making it a well-suited tool for large-scale seismic imaging tasks.

\section{Conclusions}
We introduced an DL-based optimization inverse solver to accelerate the computation of focusing functions within the Marchenko framework. More specifically, a neural network is first trained on a small subset of imaging points with pre-computed focusing functions. Once trained, the network is fixed and employed as a differentiable forward operator within an iterative optimization scheme, where BP is used to update the input and recover the up-going focusing functions.
The effectiveness of the proposed method is validated on both synthetic and field datasets from the Volve ocean-bottom acquisition. In both cases, the approach significantly reduces computational time while delivering focusing functions and final images that closely match those obtained using the conventional LSQR-based Marchenko method. Additionally, the ability of incorporating regularization and leveraging neighboring information further enhances robustness and stability during inversion. By relying exclusively on data from the target imaging area, the proposed method avoids the need for external training datasets, simplifying the workflow. Its efficiency, scalability, and ease of implementation make it a practical solution for large-scale applications.

\section{Acknowledgements}
The authors thank King Abdullah University of Science  Technology (KAUST) and the DeepWave sponsors for supporting this research. The authors also thank the members of Seismic Wave Analysis Group (SWAG) for insightful discussions. For computer time, this research used the resources of the Supercomputing Laboratory at KAUST in Thuwal, Saudi Arabia.

\section*{Conflict of Interest}
All authors have carefully reviewed and approved the manuscript for submission to the Geophysical Prospecting. The authors have no conflicts of interest to declare.

\printendnotes

% Submissions are not required to reflect the precise reference formatting of the journal (use of italics, bold etc.), however it is important that all key elements of each reference are included.
\bibliography{references}

\end{document}